\documentclass[aps,prd,twocolumn,superscriptaddress,showpacs]{revtex4} % UTOBS

\usepackage{graphicx}

\begin{document}
%\pagewiselinenumbers  % LINENO CTOBS
%\newcommand{\fix}[1]{{\bf{} }{\{\it #1}\}} %  comment this out if you don't want to see the comments
%\newcommand{\fix}[1]{} % comment this in if you want to the see the PRL without comments

\preprint{CDF/PUB/EXOTIC/CDFR/8440} %Comment this out before shipping out
\preprint{Version 3.2}

% macros
\newcommand{\ppbar}{$p\overline{p}$}
\newcommand{\ccbar}{$c\overline{c}$}
\newcommand{\bbbar}{$b\overline{b}$}
\newcommand{\ttbar}{$t\overline{t}$}
\newcommand{\MET}{\mbox{$\raisebox{.3ex}{$\not$}E_{\rm T}$}}
\newcommand{\metb}{\ensuremath{\not\!\!{E_{T}}}}
\newcommand{\gevcc}{GeV/$c^{2}$}
\newcommand{\chargino}{$\tilde{\chi}_1^{\pm}$}
\newcommand{\charginom}{$\tilde{\chi}_1^{\mp}$}
\newcommand{\neutralino}{$\tilde{\chi}_2^0$}
\newcommand{\neuone}{$\tilde{\chi}_1^0$}
\newcommand{\zvert}{$z_{vertex}$}
\newcommand{\squark}{$\tilde{q}$}
\newcommand{\gluino}{$\tilde{g}$}
\newcommand{\rp}{$R_p$}
\newcommand{\rpv}{\mbox{$\raisebox{.3ex}{$\not$}R_{p}$}}
\newcommand{\rpvb}{\ensuremath{\not\!\!{R_{p}}}}
\newcommand\pt{p_{\mathrm{T}}}
\newcommand\ET{E_{\mathrm{T}}}
\newcommand\etal{\it et al.}
\newcommand{\fix}[1]{{\bf <<< #1 !!! }}

\font\eightit=cmti8
\def\r#1{$^{#1}$}

%\begin{document}

%\begin{flushright}
%PRL Draft Version 1
%\end{flushright}
% titlepage documents

\title{Search for Anomalous Production of Multi-lepton Events 
       in \ppbar\ Collisions at $\sqrt{s}=1.96$ TeV}

%==>Comment out following for line-numbered version
\affiliation{Institute of Physics, Academia Sinica, Taipei, Taiwan 11529, Republic of China} 
\affiliation{Argonne National Laboratory, Argonne, Illinois 60439} 
\affiliation{Institut de Fisica d'Altes Energies, Universitat Autonoma de Barcelona, E-08193, Bellaterra (Barcelona), Spain} 
\affiliation{Baylor University, Waco, Texas  76798} 
\affiliation{Istituto Nazionale di Fisica Nucleare, University of Bologna, I-40127 Bologna, Italy} 
\affiliation{Brandeis University, Waltham, Massachusetts 02254} 
\affiliation{University of California, Davis, Davis, California  95616} 
\affiliation{University of California, Los Angeles, Los Angeles, California  90024} 
\affiliation{University of California, San Diego, La Jolla, California  92093} 
\affiliation{University of California, Santa Barbara, Santa Barbara, California 93106} 
\affiliation{Instituto de Fisica de Cantabria, CSIC-University of Cantabria, 39005 Santander, Spain} 
\affiliation{Carnegie Mellon University, Pittsburgh, PA  15213} 
\affiliation{Enrico Fermi Institute, University of Chicago, Chicago, Illinois 60637} 
\affiliation{Comenius University, 842 48 Bratislava, Slovakia; Institute of Experimental Physics, 040 01 Kosice, Slovakia} 
\affiliation{Joint Institute for Nuclear Research, RU-141980 Dubna, Russia} 
\affiliation{Duke University, Durham, North Carolina  27708} 
\affiliation{Fermi National Accelerator Laboratory, Batavia, Illinois 60510} 
\affiliation{University of Florida, Gainesville, Florida  32611} 
\affiliation{Laboratori Nazionali di Frascati, Istituto Nazionale di Fisica Nucleare, I-00044 Frascati, Italy} 
\affiliation{University of Geneva, CH-1211 Geneva 4, Switzerland} 
\affiliation{Glasgow University, Glasgow G12 8QQ, United Kingdom} 
\affiliation{Harvard University, Cambridge, Massachusetts 02138} 
\affiliation{Division of High Energy Physics, Department of Physics, University of Helsinki and Helsinki Institute of Physics, FIN-00014, Helsinki, Finland} 
\affiliation{University of Illinois, Urbana, Illinois 61801} 
\affiliation{The Johns Hopkins University, Baltimore, Maryland 21218} 
\affiliation{Institut f\"{u}r Experimentelle Kernphysik, Universit\"{a}t Karlsruhe, 76128 Karlsruhe, Germany} 
\affiliation{High Energy Accelerator Research Organization (KEK), Tsukuba, Ibaraki 305, Japan} 
\affiliation{Center for High Energy Physics: Kyungpook National University, Taegu 702-701, Korea; Seoul National University, Seoul 151-742, Korea; and SungKyunKwan University, Suwon 440-746, Korea} 
\affiliation{Ernest Orlando Lawrence Berkeley National Laboratory, Berkeley, California 94720} 
\affiliation{University of Liverpool, Liverpool L69 7ZE, United Kingdom} 
\affiliation{University College London, London WC1E 6BT, United Kingdom} 
\affiliation{Centro de Investigaciones Energeticas Medioambientales y Tecnologicas, E-28040 Madrid, Spain} 
\affiliation{Massachusetts Institute of Technology, Cambridge, Massachusetts  02139} 
\affiliation{Institute of Particle Physics: McGill University, Montr\'{e}al, Canada H3A~2T8; and University of Toronto, Toronto, Canada M5S~1A7} 
\affiliation{University of Michigan, Ann Arbor, Michigan 48109} 
\affiliation{Michigan State University, East Lansing, Michigan  48824} 
\affiliation{Institution for Theoretical and Experimental Physics, ITEP, Moscow 117259, Russia} 
\affiliation{University of New Mexico, Albuquerque, New Mexico 87131} 
\affiliation{Northwestern University, Evanston, Illinois  60208} 
\affiliation{The Ohio State University, Columbus, Ohio  43210} 
\affiliation{Okayama University, Okayama 700-8530, Japan} 
\affiliation{Osaka City University, Osaka 588, Japan} 
\affiliation{University of Oxford, Oxford OX1 3RH, United Kingdom} 
\affiliation{University of Padova, Istituto Nazionale di Fisica Nucleare, Sezione di Padova-Trento, I-35131 Padova, Italy} 
\affiliation{LPNHE, Universite Pierre et Marie Curie/IN2P3-CNRS, UMR7585, Paris, F-75252 France} 
\affiliation{University of Pennsylvania, Philadelphia, Pennsylvania 19104} 
\affiliation{Istituto Nazionale di Fisica Nucleare Pisa, Universities of Pisa, Siena and Scuola Normale Superiore, I-56127 Pisa, Italy} 
\affiliation{University of Pittsburgh, Pittsburgh, Pennsylvania 15260} 
\affiliation{Purdue University, West Lafayette, Indiana 47907} 
\affiliation{University of Rochester, Rochester, New York 14627} 
\affiliation{The Rockefeller University, New York, New York 10021} 
\affiliation{Istituto Nazionale di Fisica Nucleare, Sezione di Roma 1, University of Rome ``La Sapienza," I-00185 Roma, Italy} 
\affiliation{Rutgers University, Piscataway, New Jersey 08855} 
\affiliation{Texas A\&M University, College Station, Texas 77843} 
\affiliation{Istituto Nazionale di Fisica Nucleare, University of Trieste/\ Udine, Italy} 
\affiliation{University of Tsukuba, Tsukuba, Ibaraki 305, Japan} 
\affiliation{Tufts University, Medford, Massachusetts 02155} 
\affiliation{Waseda University, Tokyo 169, Japan} 
\affiliation{Wayne State University, Detroit, Michigan  48201} 
\affiliation{University of Wisconsin, Madison, Wisconsin 53706} 
\affiliation{Yale University, New Haven, Connecticut 06520} 
\author{A.~Abulencia}
\affiliation{University of Illinois, Urbana, Illinois 61801}
\author{J.~Adelman}
\affiliation{Enrico Fermi Institute, University of Chicago, Chicago, Illinois 60637}
\author{T.~Affolder}
\affiliation{University of California, Santa Barbara, Santa Barbara, California 93106}
\author{T.~Akimoto}
\affiliation{University of Tsukuba, Tsukuba, Ibaraki 305, Japan}
\author{M.G.~Albrow}
\affiliation{Fermi National Accelerator Laboratory, Batavia, Illinois 60510}
\author{D.~Ambrose}
\affiliation{Fermi National Accelerator Laboratory, Batavia, Illinois 60510}
\author{S.~Amerio}
\affiliation{University of Padova, Istituto Nazionale di Fisica Nucleare, Sezione di Padova-Trento, I-35131 Padova, Italy}
\author{D.~Amidei}
\affiliation{University of Michigan, Ann Arbor, Michigan 48109}
\author{A.~Anastassov}
\affiliation{Rutgers University, Piscataway, New Jersey 08855}
\author{K.~Anikeev}
\affiliation{Fermi National Accelerator Laboratory, Batavia, Illinois 60510}
% ...
\collaboration{CDF Collaboration\footnote{With visitors from $^a$University of Athens, 
$^b$University of Bristol, 
$^c$University Libre de Bruxelles, 
$^d$Cornell University, 
$^e$University of Cyprus, 
$^f$University of Dublin, 
$^g$University of Edinburgh, 
$^h$University of Heidelberg, 
$^i$Universidad Iberoamericana, 
$^j$University of Manchester, 
$^k$Nagasaki Institute of Applied Science, 
$^l$University de Oviedo, 
$^m$University of London, Queen Mary and Westfield College, 
$^n$Texas Tech University, 
$^o$IFIC(CSIC-Universitat de Valencia), 
}}
\noaffiliation

\date{\today}
%\fix{Draft Version 2.2}

%\doublespace

\begin{abstract}
We report a search for the anomalous production of events with
multiple charged leptons in \ppbar~collisions at $\sqrt{s} = 1.96~\rm{TeV}$
using a data sample corresponding to an integrated luminosity 
of 346~pb$^{-1}$ collected by the CDF II detector at the Fermilab Tevatron.
%We report a search for anomalous production of events with
%multiple charged leptons in \ppbar~collisions at $\sqrt{s} = 1.96~\rm{TeV}$ 
%using an integrated luminosity of 346~pb$^{-1}$ of data collected by 
%the CDF II detector at the Fermilab Tevatron.  
The search is divided into three-lepton and
four-or-more-lepton data samples.
We observe six events in the three-lepton sample and
zero events in the $\geq$4-lepton sample.
Both numbers of events are consistent with standard model background 
expectations.  Within the framework of an R-parity 
violating supergravity model, the results are interpreted as 
%limits on masses of 
mass limits on
the lightest neutralino (\neuone) and
chargino (\chargino) particles.  
For one particular choice of model parameters, the limits are 
$M$(\neuone)$>$110~GeV/$c^{2}$ and 
$M$(\chargino)$>$203~GeV/$c^{2}$ at 95\% confidence level; the variation
of these mass limits with model parameters is presented.

\end{abstract}

%==>Comment out following 2 lines in numbered version
\pacs{13.85.Qk  12.60.Jv  13.85.Rm  14.80.Ly  13.85-t}
\keywords{CDF, lepton, supersymmetry}

\maketitle

% ===========>>> Section 1: Introduction

Numerous attempts to resolve theoretical problems with the standard 
model of elementary particle physics (SM) require the existence of 
new particles with masses at the 
electroweak scale, $\sim$100~\gevcc~\cite{leptoquark,axion,msugra,susy}.
At the Tevatron and LHC hadron colliders, the
production cross sections for these particles are predicted to
be orders of magnitude smaller than for SM processes.
%a means of reducing the background while
%preserving new physics signals is required.
%a method is required to distinguish new physics signals from backgrounds.  
%At hadron colliders, such 
Analysis methods for reducing background levels while 
preserving new physics signals
include searching for leptons ($\ell$) \cite{leptons} with large 
momentum transverse to the 
beam  axis~($\pt$), like-sign leptons, or signatures with transverse energy
imbalance~(\MET).  Requiring several leptons
effectively reduces the SM backgrounds while maintaining sensitivity to low
\MET~regions.  In this Letter, we present a search for 
%non-SM 
an excess of events containing three or more leptons
above the SM prediction.  
We use a data sample with an 
integrated luminosity of 346~pb$^{-1}$  of
\ppbar~collisions at $\sqrt{s}=1.96$~TeV collected by the CDF II detector
from March~2002 to August~2004.

% ===========>>> Section 2: New physics models w/ multi-leptons
Several new physics models predict final states with four or more 
leptons that can be produced in \ppbar~collisions, including R-parity violating supersymmetry (\rpv~SUSY)
\cite{rpv_susy}, non-minimal supersymmetric models (nMSSM) \cite{nmssm}, 
and production of doubly-charged Higgs pairs \cite{h++}.
In \rpv~SUSY, the lightest supersymmetric
particle (LSP) may decay directly into two leptons and a neutrino.
With a pair of LSPs in each event, there could be at least four
leptons plus \MET.  
Another possibility is the nMSSM, which includes an
expanded Higgs sector and their supersymmetric partners.  ``Cascade" decays
of these particles may result in multiple leptons, jets, and less \MET~
than in other SUSY models.  
%Since the LSP mass limits in the nMSSM are much lower 
%than in other SUSY models, there may be only small \MET. 
Alternatively, doubly-charged Higgs ($H^{\pm\pm}$) may be 
produced in pairs in left-right symmetric models \cite{lr}. 
Under certain conditions,
the $H^{\pm\pm}$ primarily decays into two leptons, producing 
a final state of four leptons  without \MET~or jets.
In order to be sensitive to multiple new physics models, 
no \MET~or jet cuts are applied.  
The results of this search are interpreted using an \rpv~SUSY model within
an mSUGRA framework only.
A brief introduction to mSUGRA and \rpv~SUSY phenomenology is included below.
%This search focuses on \rpv~SUSY while attempting to reduce
%model dependence by not imposing \MET~or jet requirements.  
%For this reason, we discuss \rpv~SUSY phenomenology in more detail below. 

% ===========>>> Section 3: RPV phenomenology

In the minimal supersymmetric model \cite{susy}, there are two Higgs doublets,
as well as supersymmetric partners for every SM particle 
that differ by $1/2$ unit of spin.
The superpartners of the electroweak gauge 
bosons and Higgs bosons mix to produce four neutral mass 
states~($\tilde{\chi}^0_{1-4}$) and four charged 
states~($\tilde{\chi}^\pm_{1,2}$).  
%We use the supergravity framework (mSUGRA), a constrained version of the
%MSSM in which there are five free parameters: 
%the universal gaugino mass ($M_{1/2}$), the 
The supergravity framework (mSUGRA) \cite{msugra} 
has five free parameters: the universal gaugino mass ($M_{1/2}$), the 
universal scalar mass ($M_0$), the ratio of the Higgs vacuum expectation values
(tan$\beta$), the trilinear coupling ($A_0$), and the sign of the higgsino
mass parameter ($\mu$).

\begin{figure}
\centerline{\includegraphics[scale=0.5]{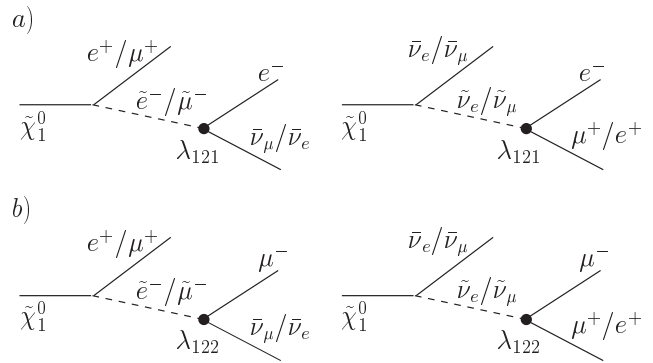}}
\caption{\label{rpv_diagram} Lowest order Feynman diagrams for \rpvb~SUSY decay
of the LSP via virtual sleptons governed by 
a) the $\lambda_{121}$ coupling, and b) the $\lambda_{122}$ coupling.
%The dots indicate the \rpvb~vertices.
}
\end{figure}

R-parity~(\rp) is a quantum number defined such that all 
SM particles have value $+1$ and all 
superpartners have value $-1$.  
%If \rp~is conserved, the LSP is stable.  
%Yet, \rp~violation is allowed under certain conditions.
Many searches assume that \rp~is conserved so that the LSP is stable,
although there are viable SUSY models with \rp~violation.
One scenario includes a single \rpv~term in the renormalizable
superpotential, $\lambda_{ijk}L_iL_j\overline{E}_k$, 
where $\lambda_{ijk}$ are lepton number violating coupling strengths; 
$i$,$j$, and $k$ are family indices ranging from 1-3;
$L_i$ is the left-handed lepton doublet superfield; and 
$\overline{E}_i$ is the right-handed charged lepton singlet superfield.  
In this scenario, superpartners are produced in pairs at
hadron colliders.  
Muon decay measurements \cite{rpv_susy} constrain $\lambda_{12k}\lesssim 0.14$
%The value of $\lambda_{12k}$ cannot be large due to muon decay measurements \cite{rpv_susy}, 
so that heavier superpartners
will cascade decay into the LSP with nearly 100\%
branching ratio, producing at least two LSPs per event.
The LSP, which is the \neuone~in mSUGRA for the values of $M_{1/2}$
pertinent to this study, 
will then decay into two leptons plus a neutrino, as shown in 
Fig.~\ref{rpv_diagram}, resulting in a final state of four or more leptons.  
The detector acceptance limits our sensitivity to 
$\lambda_{12k} \gtrsim 3\cdot10^{-3}$, since a 
smaller value will result in the LSP decaying far from the interaction region
\cite{thesis}.
%
%JH what does the following sentence mean?
%
%The maximum sensitivity is achieved when prompt
%electrons or muons are exclusively produced.  
%The value of $\lambda_{12k}$ must not be 
%so small that the LSP decays far
%from the interaction region.
%, and must not be so large that it
%violates constraints from muon decay measurements.
%The relevant ranges 
%of $\lambda_{121}$ and $\lambda_{122}$ 
%are $3\cdot10^{-3}\lesssim \lambda_{12k}\lesssim 0.14$.
%\begin{equation}
%3\cdot10^{-4}\lesssim \lambda_{12k} < 0.05 * 
%==> Use following replacement in numbered version
%3\cdot10^{-3} \lesssim \lambda_{12k} < 0.05 
%\frac{\mathrm{M}(\tilde{e},\tilde{\mu})}{100 \ {\mathrm{GeV}/c^2}} \approx 0.14.
%\label{eq:lambda_limits}
%\end{equation}    
At least two of the four leptons will be electrons (muons)
for LSP decays governed by the $\lambda_{121}$ ($\lambda_{122}$) coupling.
Final states with taus are not used since they
are less efficiently identified. 
Recently, lower mass limits have been set by the D\O~Collaboration
on the $\tilde{\chi}^0_1$ ranging from
115-119~\gevcc~and on the  
$\tilde{\chi}^\pm_1$ from 229-234~\gevcc~in this type of \rpv~SUSY model,
by searching for three leptons plus \MET~\cite{d0_rpv}.  Previously, the
limits from LEP were $\sim$53~\gevcc~on the $\tilde{\chi}^0_1$
and 103~\gevcc~on the~$\tilde{\chi}^\pm_1$~\cite{lep_limits}.
 
% ===========>>> Section 4: Detector

The components of the CDF II detector relevant to this analysis are
briefly described here; a more complete
description can be found elsewhere~\cite{CDF}.
The $\pt$ and $\eta$~\cite{cdfdefs} of charged particles  
are measured by 
%
% JH - is it 8-layer or some better way to describe??
%an eight-layer 
a
silicon strip detector~\cite{SIstrip}
and a 96-layer drift chamber (COT)~\cite{COT} inside
a $1.4$ T solenoidal magnetic field.
The COT provides coverage with high efficiency for $|\eta|<1$.
For $1<|\eta|<2$,
%At larger $|\eta|$, 
the silicon detector is mainly used.
% for measuring charged particles.
Electromagnetic (EM) and hadronic calorimeters surround the tracking system.
They are segmented in a projective tower geometry and measure energies
of charged and neutral particles in the central
($|\eta|<1$) and end-plug ($1.1<|\eta|<3.6$) regions.
Each calorimeter has an EM shower profile detector positioned
at the shower maximum.
Four layers of planar drift chambers located outside the central hadron
calorimeters and another set behind a 60~cm thick steel absorber
detect muons with $|\eta|<0.6$.  Additional drift chambers and 
scintillation counters detect muons in the region $0.6<|\eta|<1.0$.
Gas Cherenkov counters~\cite{CLC} measure the average number of
inelastic \ppbar~collisions per bunch crossing and thereby determine
the luminosity.

% ===========>>> Section 5: Lepton identification

Electron identification criteria include a reconstructed track 
associated with a cluster of energy in the EM calorimeter for $|\eta|<2.0$.
``Tight" central electrons and end-plug electrons are required to 
have high track quality, 
calorimeter cluster $E_T$ that is consistent with the track $\pt$, 
a high fraction (usually 95\%) of the total calorimeter energy 
deposition in the EM section, 
and an EM lateral shower profile 
%similar to expectations from 
consistent with test beam measurements. 
A ``loose" central electron sample 
is also collected by removing the shower profile requirements.
Muons are identified for $|\eta|<1$ by a charged track consistent 
with being minimum ionizing in the calorimeters matched to a 
reconstructed track segment (``stub") in one of the muon drift chambers.
%Muons are identified  for $|\eta|<1$ by a track matched to a 
%reconstructed muon track segment (``stub") in one of the muon drift chambers.
%Muon selection criteria include associated calorimeter energy
%consistent with a minimum ionizing particle and quality requirements 
%on the track and muon stub.  
Stubless muons are also accepted for $|\eta|<1$ if all other
criteria are satisfied except for the stub requirement and
the track does not point to any of the muon drift chambers.
All tracks associated with electrons or muons must be consistent 
with being produced at the \ppbar~collision point. 
%For both electrons and muons, the tracks must be consistent  
%with production at the \ppbar~collision point. 
Leptons are required to be separated from each other by 
$\Delta R = \sqrt{(\Delta\eta)^2 + (\Delta\phi)^2} > 0.4$ 
and to have $|z|<60$~cm at the point of closest approach to the primary vertex.
Leptons are required to be isolated such that $\ET^{iso}$ divided 
by the $\ET$ of the electron or $\pt$ of the muon is less than $0.1$, where 
$\ET^{iso}$ is the transverse energy within a cone of $\Delta R < 0.4$ 
that is unassociated with the lepton.  
%Prompt leptons are defined such that the impact parameter is 
%$\lesssim$ 0.02 cm from the \ppbar~collision point.

% ===========>>> Section 6: Event selection

The data were collected with lepton triggers~%\cite{cdf2}
requiring at least one central electron (muon) candidate with 
$\ET > 18$~GeV ($\pt > 18$~GeV/$c$). 
We select events with two or more
leptons that have $\ET$ ($\pt$) $>$ 20, 8, and 5~GeV~(GeV/$c$)
for the first, second, and additional leptons, respectively.
%when ranked in order
%of $\ET$ ($\pt$).  The first lepton must be central.
%These kinematic cuts accept a large fraction of the \rpv~SUSY events.
At least one identified lepton must pass the trigger requirements.
From this sample, 
%the dilepton data sample
trilepton and $\geq$4-lepton events 
that pass additional cuts (described below) are used as signal candidates,  
while dilepton events are used to validate the background prediction.

% ===========>>> Section 6: Backgrounds and RpV SUSY signal estimation

%The event selection criteria are determined by evaluating
%cross sections and detector acceptances for 
%sources of SM background as well as for
%hypothetical \rpv~SUSY signals, largely using Monte Carlo (MC) simulation
%but with corrections derived from CDF data.
%The numbers of simulated events are normalized to 
%$\int\mathcal{L}\ dt$~=~346~pb$^{-1}$.
%The acceptances are corrected for 
%$\epsilon_{data}/\epsilon_{MC}$, where $\epsilon_{data}$ ($\epsilon_{MC}$) is 
%the efficiency for lepton reconstruction, lepton identification,
%and photon conversion removal requirements measured in data (MC).
%This correction varies from $\sim$0.6-1.0, depending primarily on the number
%and types of identified leptons in the event.  
%Lepton reconstruction and identification efficiencies are measured from DY and 
%$J/\psi \rightarrow \ell\ell$ events in data and MC simulation.  
%Trigger efficiencies are
%determined from $W \rightarrow e\nu$ events for electrons and 
%$Z\rightarrow \mu\mu$ events for muons.

Several SM processes result in final states with three or more isolated
leptons.  The dominant contributions include dileptons from the
Drell-Yan $Z/\gamma^*$ process (DY) with additional 
leptons from photon conversions or misidentified jets, and 
secondary contributions from diboson 
($WZ$, $ZZ$)
%($WZ/\gamma^*$, $Z/\gamma^*~Z/\gamma^*$)
production.  Except for those due to misidentified jets, the backgrounds
are estimated using Monte Carlo (MC) simulation.
%The numbers of simulated events are normalized to 
%$\int\mathcal{L}\ dt$~=~346~pb$^{-1}$.
The simulated detector acceptances are not a 100\% accurate
representation of the data and therefore are corrected by  
$\epsilon_{\rm data}/\epsilon_{\rm MC}$, 
where $\epsilon_{\rm data}$ ($\epsilon_{\rm MC}$) is 
the efficiency for lepton reconstruction, lepton identification,
and photon conversion removal requirements measured in data (MC).
This correction varies from 0.6-1.0 per event, depending on the number
and types of identified leptons.  
Lepton reconstruction and identification efficiencies are measured from DY and 
$J/\psi$ dilepton events in data and MC simulation.  
Trigger efficiencies are
determined from $W \rightarrow e\nu$ events for electrons and 
from $Z\rightarrow \mu\mu$ events for muons.  
The probability for a jet to be mis-identified as a lepton is determined 
as a function of $\ET$ for electrons and $\pt$ for muons
from jet data samples, by counting the numbers
of jets that remain after applying the lepton identification requirements.
A similar procedure is described in more detail in \cite{fake_rate}.  

Samples of \rpv~SUSY simulated events are generated using 
\begin{sc}pythia\end{sc} 
% 6.216 
\cite{pythia}
within the mSUGRA framework with input of the supersymmetric couplings and
particle masses from \begin{sc}isajet\end{sc} 
% 7.51 
\cite{isajet}.  
For each point in the SUSY parameter space, the next-to-leading-order 
cross section is calculated with \begin{sc}prospino\end{sc}2 
\cite{prospino}.  In all
cases $\tilde{\chi}^0_2 \tilde{\chi}^{\pm}_1$ 
and $\tilde{\chi}^{\pm}_1 \tilde{\chi}^{\mp}_1$ production
dominate and are the only processes generated.
The \neuone~mass, which is roughly 
proportional to $M_{1/2}$, has the largest impact
on the signal acceptance.  The \neuone~mass is also affected by 
the sign($\mu$), but has little dependence on $M_0$, tan$\beta$, and $A_0$.
Therefore, three of the five mSUGRA parameters are held constant:
$M_0$~=~250~GeV/$c^2$, tan$\beta = 5$, and $A_0 = 0$;
while sign($\mu$) and $M_{1/2}$ are allowed to vary.
Since the \neuone~mass decreases with larger values of $M_0$ and tan$\beta$,
the chosen values are expected to produce conservative results.

% Should something be said about charged lepton production in cascade 
% decays as a function of the parameter space? -A.A.

%Of the five mSUGRA parameters, three
%are held constant: $M_0$~=~250~GeV/$c^2$, tan$\beta = 5$, 
%and $A_0 = 0$;
%while sign($\mu$) and $M_{1/2}$ are allowed to vary.  

%For the defined signal regions,
%the major sources of SM background are expected to come from Drell-Yan
%production plus jets where one or more jets are misidentified as leptons,
%and from Drell-Yan plus $\gamma \rightarrow e^+e^-$ conversions.  

%%%%%%%%%%%%%%%%%%
%\begin{table}[htbp]
%\begin{center}
%\begin{tabular}{llccc}
%\hline
%Electron & Type & Tight & $|\eta|$ & \\
%\hline
%	 & TCE & yes & $<$ 1.1 & \\
%	 & LCE & no  & $<$ 1.1 & \\
%	 & Plug & no & 1.2 - 2.0 & \\
%\hline
%Muon     & Type & Tight & $|\eta|$ & Muon Stub \\
%\hline
%	 & CMUP & yes & $<$ 0.6 & yes \\
%	 & CMX	& yes & 0.6 - 1.1 & yes \\
%	 & CMIO & no  & $<$ 1.1 & no \\
%\hline
%\end{tabular}
%\end{center}	 
%\caption{Lepton types.}
%\label{tbl:leptons}
%\end{table}
%%%%%%%%%%%%%%%%%%

To reduce SM background contamination, an event is removed 
if any of the following conditions are met 
for opposite-sign, same-flavor lepton pairs:  
invariant mass between 76-106~GeV/$c^2$ (``$Z$ veto''), 
invariant mass below 15~GeV/$c^2$, or
$160^{\circ}<\Delta\phi<200^{\circ}$.  
Identified cosmic ray events are removed.
Electrons are removed if there is a partner track identified as coming 
from a photon conversion.  
An event is also rejected if the invariant mass 
of the two highest transverse energy leptons is below 20~GeV/$c^2$ 
to reduce heavy flavor and radiative photon conversion backgrounds.
In order to improve signal-to-background in the 3-lepton data sample only, 
we require that one of the two leading leptons be an electron
when evaluating the
$\lambda_{121}$ scenario, and a muon for the
$\lambda_{122}$ scenario.  This definition leads to a 22\% overlap
for the background and $\sim$15\% overlap for the signal (depending on
model parameters) between the trilepton event samples.
%The latter requirement is not used for the nearly background-free
%$\geq$4-lepton data sample.
After all selection criteria, the signal acceptances
%times $\epsilon_{data}/\epsilon_{MC}$ 
are $\sim$11\% and $\sim$4\% for the trilepton and 
$\geq$4-lepton data samples, respectively.

% ===========>>> Section 7: Uncertainty

Uncertainties are determined separately for SM background 
and \rpv~SUSY expectations, in each data sample. 
The sources of systematic uncertainty on the trilepton background prediction
include jets misidentified as leptons (12-13\%), lepton identification
efficiency (5-6\%), luminosity measurement (6\%), choice of parton
distribution functions (2\%), cross section (5-6\%), 
photon conversion identification (3\%), and initial state
radiation (4\%).  For events with $\geq$4-leptons,
the uncertainty on the background is dominated by jets misidentified as leptons
(41\%) and $Z/\gamma^*$ + $\gamma$ MC statistics (38\%).  The total systematic 
uncertainty on the signal acceptance ranges from 12-15\%, based on 
the number and types of leptons, where the largest contribution
is due to the uncertainty on the low-$\pt$~($<20$~GeV/$c$) 
lepton identification efficiency. 

%The following sources of systematic uncertainty are considered:
%jets mis-identified as leptons, 
%lepton identification efficiency,
%photon conversion identification efficiency, 
%luminosity measurement, choice of parton distribution
%functions, cross section, and initial state radiation.
%Uncertainties are determined separately for SM background expectations 
%and \rpv~SUSY in each data sample.  
%For the background, the 
%dominant uncertainties are due to  statistics in the $Z/\gamma^*$ + $\gamma$ 
%MC samples and the fake rates.  For the \rpv~SUSY expectation,
%the largest uncertainty is due to the low-$\pt$~($<20$~GeV/$c$) 
%lepton identification efficiency.   

% ===========>>> Section 8: Validation

The background prediction is validated through the use of 
control regions in which one or more of the
event selection criteria are inverted.  
For each control region, the SM prediction is compared to
the data, as shown in Table \ref{tbl:crs}.
In the $ee$ and $\mu\mu$ dilepton control regions, the 
background is dominated by DY.  In the case of $e\mu$ control 
regions, the largest background is $Z/\gamma^* \rightarrow \tau\tau$;
however, there  are also significant contributions from $WW$, $t\overline{t}$,
and jets misidentified as leptons.  
For trilepton control regions, the largest backgrounds 
originate from DY events where the third lepton is due to
either a photon conversion or a misidentified jet.
%is either 
%a photon conversion or a jet misidentified as a lepton.  
%An example of a
%distribution in the trilepton control region is shown in Fig.
%\ref{trilepton_met}.
The agreement between predicted and observed events in the
control regions indicates that the backgrounds
are validated, and we proceed to examine the signal data samples.

% ===========>>> Section 9: Table combining control and signal region #s

\begin{table}
\begin{center}
\caption{Predicted and observed events found in the control and signal samples.  
The leptons are listed in order of decreasing transverse energy.
The first lepton
must either be a tight central electron or ``stub muon."
Systematic and statistical uncertainties are added in quadrature. 
%Notation:  X = third or fourth lepton (having lowest $p_T$),
%TCE = tight central electron, StMu = stubbed muon, 
%$Z$ = invariant mass within $Z$ window from 76-106~\gevcc~, 
%No $Z$ = invariant mass outside $Z$ window. 
\label{tbl:crs}}
\begin{tabular}{llll}
\hline
\hline
\bf Region & \bf Criteria Failed & \bf Predicted & \bf Observed\\
           &                     & \bf Background & \bf Events\\
\hline
%\hline
\multicolumn{4}{l}{\bf Two-Lepton Control Samples} \\
\hline
$ee$ & $Z$ veto 	& $13948 \pm 1536$ 	& 14019 \\
$ee$ &			& $2142  \pm 230$	& 2125 \\
$\mu\mu$ & $Z$ veto 	& $7474  \pm 809$	& 7499 \\
$\mu\mu$ & 		& $1264  \pm 141$	& 1339 \\
$e\mu$ & 		& $117.6 \pm 12.9$	& 112 \\
$\mu e$ & 		& $186.8 \pm 22.9$	& 203 \\
\hline
%\hline
\multicolumn{4}{l}{\bf Three-Lepton Control Samples} \\
\hline
$ee\ell$       	& $Z$ veto       & $8.8  \pm 1.9$  & 12 \\
$\mu\mu\ell$   	& $Z$ veto       & $5.1  \pm 1.2$  & 2\\
$e\mu\ell$     	& $Z$ veto       & $0.55 \pm 0.04$ & 0\\
$\ell\ell\ell$  & $Z$ veto       & $14.4 \pm 2.9$  & 14 \\
$ee\ell$       	& $\Delta\phi$ only & $2.1  \pm 0.3$  & 2\\
$\mu\mu\ell$   	& $\Delta\phi$ only & $1.2  \pm 0.2$  & 4\\
$e\mu\ell$     	& $\Delta\phi$ only & $0.35 \pm 0.04$ & 0\\
$\ell\ell\ell$  & $\Delta\phi$ only & $3.7  \pm 0.3$  & 6 \\
\hline
%\hline
\multicolumn{4}{l}{\bf Four-or-More-Lepton Control Samples} \\
\hline
$\ell\ell\ell\ell$ & $Z$ veto       & $0.15  \pm 0.02 $ & 0\\
$\ell\ell\ell\ell$ & $\Delta\phi$ only & $0.006 \pm 0.003$ & 0\\
\hline
%\hline
\multicolumn{4}{l}{\bf Three-Lepton Signal Samples} \\
\hline
$\lambda_{121}$ scenario & & $3.1 \pm 0.8$ & 5 \\
$\lambda_{122}$ scenario & & $1.9 \pm 1.0$ & 1 \\
%\hline
\hline
\multicolumn{4}{l}{\bf Four-or-More-Lepton Signal Sample} \\
\hline
$\lambda_{121}$, $\lambda_{122}$ scenarios& & $0.008 \pm 0.004$ & 0 \\
\hline
\hline
\end{tabular}
\end{center}
\end{table}

\begin{figure}
%\centerline{\includegraphics[scale=0.42]{AllSigReg_met_bgerr.eps}}
\centerline{\includegraphics[scale=0.45]{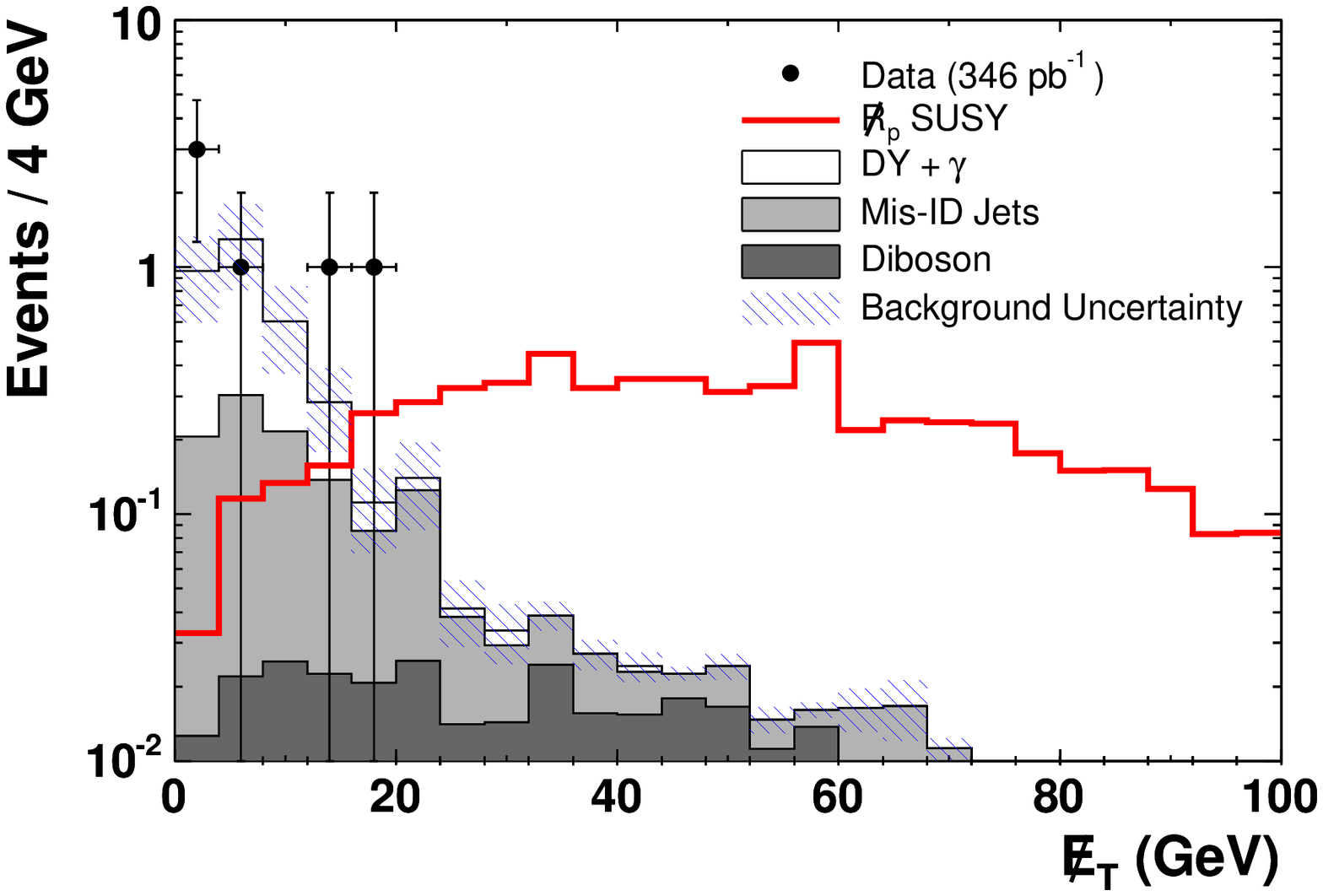}}
\caption{\label{trilepton_met} 
\metb~for events in the trilepton signal samples. For comparison, an \rpvb~SUSY sample is shown with $M$(\neuone) = 99 \gevcc.   }
\end{figure}

% ===========>>> Section 10: Results

\begin{figure}
\centerline{\includegraphics[scale=0.43]{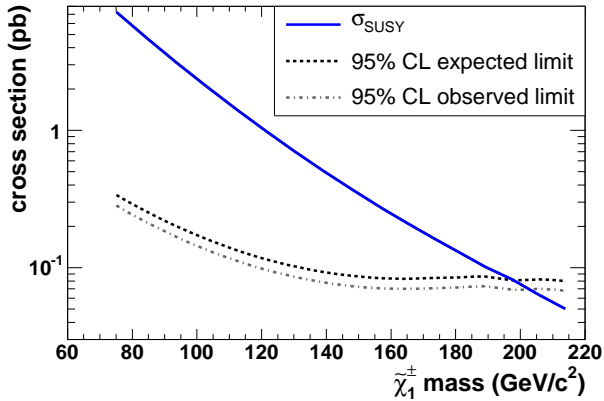}}
\caption{\label{cross_vs_mass} 
Experimental 95\% C.L. cross section limits and~theoretical~\rpvb~SUSY
 cross sections versus LSP mass, 
for $3\cdot 10^{-3} \lesssim \lambda_{122} <$ 0.14,  
$M_{0}$ = 250 \gevcc, tan$\beta$ = 5, $A_0$ = 0, and $\mu > 0$.}
%\rpv~theoretical cross section and experimental cross section limits versus LSP mass.}
\end{figure}

In the trilepton data samples, 
a total of 6 events are observed:
5 events with an expected background of 3.1~$\pm$~0.7~(stat.)~$\pm$~0.4~(syst.)
events for the $\lambda_{121}$ scenario, 
and one event with an expected 
background of 1.9~$\pm$~1.0~(stat.) $\pm$~0.3~(syst.) events for the 
$\lambda_{122}$ scenario. 
The \MET~distribution of the observed events are consistent with 
the background prediction, as shown in Fig. \ref{trilepton_met}.
No events are observed in the $\geq$4-lepton 
data sample, with an expected background of 
0.008~$\pm$~0.003~(stat.)~$\pm$~0.003~(syst.) events.  
%The observed trilepton events are compared to the 
%background expectation in Fig. \ref{trilepton_met}.
We interpret the results as being consistent with the hypothesis of no 
signal and therefore set mass limits on the \neuone~and \chargino~particles.  
The cross section limits 
%as a function of the superpartner masses
are calculated by combining the 
3-lepton and $\geq$4-lepton signal samples using a multi-channel Bayesian 
method similar to \cite{bayes}, shown for one 
\rpv~SUSY scenario in Fig. \ref{cross_vs_mass}.  The limit calculation 
takes into account correlations 
%treats uncertainties as nuisance parameters that are allowed to be
%correlated 
between the 3-lepton and $\geq$4-lepton samples.
The resulting 
%$\tilde{\chi}_1^0$ and$\tilde{\chi}_1^\pm$
mass limits, at 95\% confidence level (C.L.),
are presented in Table \ref{tbl:mass_limits}.
%and Fig. \ref{mass_limits}.

\begin{table}
\begin{center}
\caption{Expected and observed 95\% C.L. limits on the $\tilde{\chi}^0_1$ and
$\tilde{\chi}^\pm_1$ masses for $3\cdot 10^{-3} \lesssim \lambda_{12k} <$ 0.14, $M_0$ = 250~\gevcc, tan$\beta$~=~5, and $A_0$ = 0.}
\begin{tabular*}{0.47\textwidth}{@{\extracolsep{\fill}}lcccc}
\hline
\hline
\bf \rpvb~SUSY& \multicolumn{2}{c}{\bf M($\tilde{\chi}^0_1$) [GeV/$c^2$]} & \multicolumn{2}{c}{\bf M($\tilde{\chi}^\pm_1$) [GeV/$c^2$]} \\
\bf scenario& \bf exp. & \bf obs. & \bf exp. & \bf obs. \\
\hline
$\lambda_{121}$, $\mu > 0$ & 105 & 102 & 192 & 185 \\
$\lambda_{121}$, $\mu < 0$ & 101 &  98 & 192 & 186 \\
$\lambda_{122}$, $\mu > 0$ & 108 & 110 & 198 & 203 \\
$\lambda_{122}$, $\mu < 0$ & 103 & 106 & 195 & 202 \\
\hline
\hline
\end{tabular*}
\label{tbl:mass_limits}
\end{center}
\end{table}

%\begin{figure}
%\includegraphics[scale=0.4]{limits.eps}
%\caption{\label{mass_limits} 
%mSUGRA mass limits on the lightest 
%neutralino and chargino particles
%for various choices of \rpv~coupling type and sign of $\mu$ parameter.}
%\end{figure}

% ===========>>> Section 11: Conclusions

In conclusion, we have performed a search
for anomalous production of events with three or more
leptons using a sample of CDF Run II data
%
% JH suggested by Spokes reading group:
corresponding to 
%$\int\mathcal{L}\ dt$ = 
346 pb$^{-1}$ of integrated luminosity.  
Finding results consistent with the SM, we
set limits on the \neuone~and \chargino~masses within an
mSUGRA framework.   The \neuone~mass limits range from 
98 to 110~GeV/$c^{2}$, while the chargino mass limits
range from 185 to 203~GeV/$c^{2}$ at 95\% C.L.
depending on the choice of model parameters.  These results 
significantly improve upon the LEP limits \cite{lep_limits} 
%\cite{lep_limits} of $\sim$53 \gevcc~on the 
%\neuone~mass and 103~\gevcc~on the \chargino~mass 
and are comparable to the D\O~limits \cite{d0_rpv}.
%\cite{d0_rpv} ranging from 115-119~\gevcc~on the 
%\neuone~mass and from 229-234 on the \chargino~mass.
While no evidence for new physics was found, the $\geq$4-lepton data 
sample is virtually background free and provides 
an excellent technique for detecting
new physics with more luminosity in the future.
 
%
%As CDF Run II is expected to collect
%4-8 fb$^{-1}$ by 2009,  we look forward to repeating this
%analysis with more data.
%Additionally, in the \rpv~SUSY interpretation, 
%a looser lepton isolation requirement would decrease the model dependence,
%and looser vertex criteria would increase sensitivity to lower $\lambda$ values.

%
% Acknowledgments
%    

We thank the Fermilab staff and the technical staffs of the participating institutions for their vital contributions. This work was supported by the U.S. Department of Energy and National Science Foundation; the Italian Istituto Nazionale di Fisica Nucleare; the Ministry of Education, Culture, Sports, Science and Technology of Japan; the Natural Sciences and Engineering Research Council of Canada; the National Science Council of the Republic of China; the Swiss National Science Foundation; the A.P. Sloan Foundation; the Bundesministerium f\"ur Bildung und Forschung, Germany; the Korean Science and Engineering Foundation and the Korean Research Foundation; the Particle Physics and Astronomy Research Council and the Royal Society, UK; the Russian Foundation for Basic Research; the Comisi\'on Interministerial de Ciencia y Tecnolog\'{\i}a, Spain; in part by the European Community's Human Potential Programme under contract HPRN-CT-2002-00292; and the Academy of Finland.

%\clearpage
%
%  References
%

%%%%%%%%%%%%%%%%%%%%%%%%%%%% References section %%%%%%%%%%%%%%%%%%%%%%%%%%
% A useful Journal macro
\def\Journal#1#2#3#4{{#1} {\bf #2}, #3 (#4)}
% Some useful journal names
\def\NCA{Nuovo Cimento}
\def\NIM{Nucl. Instrum. Methods}
\def\NIMA{{Nucl. Instrum. Methods} A}
\def\NP{Nucl. Phys.} 
\def\NPB{{Nucl. Phys.} B}
\def\PLB{{Phys. Lett.}  B}
\def\PRL{Phys. Rev. Lett.}
\def\RPP{Rep. Prog. Phys.}
\def\PRD{{Phys. Rev.} D}
\def\PR{Phys. Rep.}
\def\PRP{Prog. Theor. Phys}
\def\ZPC{{Z. Phys.} C}
\def\MPL{{Mod. Phys. Lett.} A}
\def\EPJC{{Eur. Phys. J.} C}
\def\CPC{Comput. Phys. Commun.}

\renewcommand{\baselinestretch}{1}


\begin{thebibliography}{99}
\bibitem{cdfdefs}
   In the CDF geometry, $\theta$ is the polar angle with respect to the
   proton beam axis (positive $z$ direction), and $\phi$ is the 
   azimuthal angle.
   The pseudorapidity is $\eta=-\ln [ \tan(\theta/2) ]$.  
   %The transverse momentum, $\pt$, is the component of the momentum projected
   %onto the plane perpendicular to the beam axis.      
   The transverse energy, $\ET$, of a shower or calorimeter tower is 
   $E \sin\theta$, where $E$ is the energy deposited.
   The missing transverse energy is defined by \MET~$= |\sum_{i}
   E_T^i\hat{n}_i|$, where $\hat{n}_i$ is the transverse component of the
   unit vector pointing from the interaction point to calorimeter tower $i$.
\end{thebibliography}
\end{document}